\documentstyle[11pt,epsfig]{article}
\def\be{\begin{eqnarray}}
\def\ee{\end{eqnarray}}
\def\ben{\begin{enumerate}}\def\bitem{\begin{itemize}}
\def\een{\end{enumerate}}\def\eitem{\end{itemize}}
\def\bi{\bibitem}

\def\Kp{$K^+$}\def\Km{$K^-$}
\def\wkm{\omega_{\tiny{K^-}}^\star}
\def\calL{{\cal L}}

\def\prl{Phys. Rev. Lett.}\def\pr{Phys. Rev.}\def\np{Nucl. Phys.}
\def\pl{Phys. Lett.}

\def\del{\partial}

\def\roughly#1{\mathrel{\raise.3ex\hbox{$#1$\kern-.75em%
\lower1ex\hbox{$\sim$}}}}\def\lsim{\roughly<}
\def\gsim{\roughly>}

\def\itt{\indent\indent}

\def\Bp{\mbox{\boldmath$ p$}}

\def\A0{A_0}
\def\bq{\begin{equation}}
\def\eq{\end{equation}}

\renewcommand{\thefootnote}{\fnsymbol{footnote}}
\begin{document}
\begin{titlepage}
\begin{center}
%\hfill {First draft}

 \vskip 2cm {\Large \bf Strangeness Equilibration at GSI Energies}
  \vskip 2cm
   {{\large G.E. Brown$^{(a)}$, Mannque Rho$^{(b)}$} and Chaejun Song$^{(a)}$ }
 \vskip 0.2cm

{\it (a) Department of Physics and Astronomy, State University of
New York}

{\it Stony Brook, NY 11794, USA}

 {\it (b) Service de Physique Th\'eorique, CE Saclay, 91191
Gif-sur-Yvette, France}

\end{center}

\vskip 1cm

\centerline{(\today)}
 \vskip 2cm

\centerline{\bf Abstract}
 \vskip 0.5cm
We develop the notion of ``broad-band equilibration" in heavy-ion
processes involving dense medium. Given density-dependent
\Km-masses we show that the equilibration at GSI energies claimed
to hold in previous treatments down to $\sim \rho_0/4$, can be
replaced by a broad-band equilibration in which the \Km-meson and
hyperons are produced in an essentially constant ratio
independent of density. There are experimental indications that
this also holds for AGS energies. We then proceed to argue that
{\it both} $K^+$ and $K^-$ must get lighter in dense medium at
some density $\rho >\rho_0$ due to the decoupling of the vector
mesons.  As a consequence, kaon condensation in compact stars
could take place {\it before} chiral restoration since the sum of
bare quark masses in the kaon should lie below $\mu_e$. Another
consequence of the decoupling vector interactions is that the
quasi-particle picture involving (quasi)quarks, presumably
ineffective at low densities, becomes more appropriate at higher
densities as chiral restoration is approached.

\end{titlepage}
\newpage

\renewcommand{\thefootnote}{\arabic{footnote}}
\setcounter{footnote}{0}

\section{Introduction}
\itt
 Following work by Hagedorn~\cite{hagedorn} on production of
 anti-$^3$He, Cleymans et al~\cite{cleymans} have shown that for low
 temperatures, such as found in systems produced at GSI,
 strangeness production is strongly suppressed. The abundance of
 $K^+$ mesons, in systems assumed to be equilibrated, is given by
 \cite{cleymans2},
 \be
 n_{\tiny{K^+}}\sim e^{-E_{\tiny{K^+}}/T}\ \
 V\left\{g_{\bar K}\int\frac{d^3p}{(2\pi)^3}
 e^{-E_{\bar K}/T}+g_\Lambda\int \frac{d^3p}{(2\pi)^3}
 e^{-(E_\Lambda-\mu_B)/T}\right\}.\label{K+}
 \ee
 Here the $g$'s are the degeneracies. Because strangeness must be
 conserved in the interaction volume $V$, assumed to be that of
 the equilibrated system for each $K^+$ which is produced, a
particle of ``negative strangeness"~\footnote{By ``negative
strangeness" we are referring to the negatively charged strange
quark flavor. The positively charged anti-strange quark will be
referred to as ``positive strangeness."} containing $\bar{s}$,
say, $\bar K$ or $\Lambda$, must also be produced, bringing in the
$\bar K$ or $\Lambda$ phase space and Boltzmann factors. The $K^+$
production is very small at GSI energies because of the low
temperatures which give small Boltzmann factors for the $\bar K$
and $\Lambda$. Note the linear dependence on interaction volume
which follows from the necessity to include $\bar K$ or $\Lambda$
phase space.

In an extensive and careful analysis, Cleymans, Oeschler and
Redlich~\cite{cleymans2} show that measured particle multiplicity
ratios $\pi^+/p$, $\pi^-/\pi^+$, $d/p$, $K^+/\pi^+$, and
$K^+/K^-$ -- but not $\eta/\pi^0$ -- in central Au-Au and Ni-Ni
collisions at (0.8-2.0)A GeV are explained in terms of a thermal
model with a common freeze-out temperature and chemical
potential, if collective flow is included in the description. In
other words, a scenario in which the kaons and anti-kaons are
equilibrated appears to work well. This result is puzzling in
view of a recent study by Bratkovskaya et al~\cite{brat} that
shows that the $K^+$ mesons in the energy range considered would
take a time of $\sim 40$ fm/c to equilibrate. We remark that this
is roughly consistent with the estimate for higher energies in the
classic paper by Koch, M\"uller and Rafelski~\cite{koch} that
strangeness equilibration should take $\sim 80$ fm/c. Such
estimates have been applied at CERN energies and the fact that
emergent particle abundances are described by Boltzmann factors
with a common temperature $\sim 170-180$ MeV has been used as
part of an argument that the quark/gluon plasma has been observed.

We interpret the result of \cite{cleymans2} as follows. Since
free-space masses are used for the hadrons involved, Cleymans et
al~\cite{cleymans2} are forced to employ a $\mu_B$ substantially
less than the nucleon mass $m_N$ in order to cut down $\Lambda$
production as compared with $K^-$ production, the sum of the two
being equal to $K^+$ production. This brings them to a diffuse
system with density of only $\sim \rho_0/4$ at chemical
freeze-out. But this is much too low a density for equilibration.

We shall first show how this situation can be improved by
%reducing the $K^-$ mass to $m_K^\star < m_K$.
replacing the $K^-$ mass by the \Km energy at rest $\wkm
\equiv\omega_-(k=0) < m_K$. (The explicit formula for
$\omega_\pm$ is given later, see eq. (\ref{energy}).) In doing
this, we first have to reproduce the $K^+$ to $K^-$ ratio found
in the Ni + Ni experiments~\cite{menzel}:
 \be
n_{K^+}/n_{K^-}\simeq 30.\label{ratio}
 \ee
Cleymans et al reproduce the earlier smaller ratio of $21\pm 9$
with $\mu_B=750$ MeV and $T=70$ MeV. How this or rather
(\ref{ratio}) comes out is easy to see. The ratio of the second
term on the RHS of eq.(\ref{K+}) to the first term is roughly the
ratio of the exponential factors multiplied by the phase space
volume
 \be
R=\frac{g_\Lambda}{g_{\bar K}}
\left(\frac{\bar p_\Lambda}{\bar p_{\bar K}}\right)^3
\frac{e^{-(E_\Lambda-\mu_B)/T}}{e^{-E_{\tiny{K^-}}/T}}\approx
\left(\frac{m_\Lambda}{m_{\bar K}}\right)^{3/2}
\frac{e^{-(m_\Lambda-\mu_B)/T}}{e^{-m_{\tiny{K^-}}/T}}\approx 21
 \label{R} \ee
where we have used $g_\Lambda\approx g_{\bar K}\simeq 2$, $(\bar
p)^2/2 m\simeq \frac 32 T$, $E_\Lambda=1115$ MeV and
$E_{\tiny{\bar K}}=495$ MeV. We are able to convert $E_\Lambda$
to $m_\Lambda$ and $E_{\tiny{K^-}}$ to $m_{\tiny{K^-}}$ because
$E_\Lambda\simeq m_\Lambda + \frac 32 T$ and
$E_{\tiny{K^-}}\simeq m_{\tiny{K^-}}+ \frac 32 T$ and the thermal
energies cancel out in the ratio. This works as long as the
masses are more than $\sim 3T$, where the nonrelativistic
approximation is valid. Inclusion of the $\Sigma$ and $\Xi$
hyperons would roughly increase this number by 50\% with the
result that the ratio of $K^-$ to $\Lambda$, $\Sigma$, $\Xi$
production is~\footnote{In order to reproduce this result with
$\mu_B=750$ MeV and $T=70$ Mev within our approximation, we have
assumed only the $\Sigma^-$ and $\Sigma^0$ hyperons to
equilibrate with the $\Lambda$. This may be correct because the
$\Sigma^+$ and $\Xi$ couple more weakly. Inclusion of the latter
could change our result slightly. Probably they should be
included in analysis of the AGS experiments at higher energies
where they would be more copiously produced.}
 \be
\frac{n_{K^-}}{n_{\Lambda+ \Sigma+ \Xi}}\simeq 1/32.
 \ee
Since a $K^+$ must be produced to accompany each particle of one
unit of strangeness (to conserve strangeness flavor), we then have
 \be
n_{K^+}/n_{K^-}\sim 33.
 \ee
This is consistent with the empirical ratio (\ref{ratio}). It
should be noted that had we set $\mu_B$ equal to $m_N$, we would
have had the $K^+$ to $K^-$ ratio to be $\sim 280$ because it
costs so much less energy to make a $\Lambda$ (or $\Sigma$)
rather than $K^-$ in this case. In other words the chemical
potential $\mu_B$ is forced to lie well below $m_N$ in order to
penalize the hyperon production relative to that of the $K^-$'s.

One can see from fig.5 in Li and Brown~\cite{LB} that without
medium effects in the $K^-$ mass, the $K^+/K^-$ ratio is $\sim
100$, whereas the medium effect decreases the ratio to about 23.
This suggests how to correctly redo the Cleymans et al's analysis,
namely, by introducing the dropping $K^-$ mass into it.

In Appendix A we show that positive strangeness production takes
place chiefly at densities greater than $2\rho_0$. As the
fireball expands to lower densities the amount of positive
strangeness remains roughly constant. The number of \Kp's is such
as to be in equilibrium ratio $K^+/\pi^+$ with the equilibrated
number density of pions at $T=70$ MeV, $n_\pi\approx 0.37\,
T_{197}^3\, {\rm fm}^{-3}$. Only in this sense do the \Kp's
equilibrate.

It is amusing to note that the ``equilibrated ratio" of $\sim 30$
for the $n_{K^+}/n_{K^-}$ holds over a large range of densities
for $T=70$ MeV, once density-dependent \Km masses are introduced,
in that the ratio $R$ of (\ref{R}) is insensitive to density.
(Remember that because of the small number of \Km's, the number
of \Kp's must be nearly equal to the number of hyperons,
$\Lambda$, $\Sigma^-$ and $\Xi$, in order to conserve
strangeness.) This insensitivity results because $\wkm $
decreases with density at roughly the same rate as $\mu_B$
increases. We can write $R$ of (\ref{R}), neglecting possible
changes in $T$ and $m_{\bar{K}}$ in our lowest approximation, as
 \be
R=\left(\frac{m_\Lambda}{m_{\bar{K}}}\right)^{3/2} e^{(\mu_B+\wkm
)/T}\, e^{-m_\Lambda/T}.\label{Rp}
 \ee
As will be further stressed later, the most important point in
our arguments is that {\it the $\mu_B+\wkm $ is
nearly constant with density}. This is because whereas $\mu_B$
increases from 860 MeV to 905 MeV as $\rho$ goes
%from $\rho_0$ to $2\rho_0$, $m_K^\star $ decreases from 415 MeV to 350 MeV,
from $1.2\rho_0$ to $2.1\rho_0$, $\wkm$ decreases from 370 MeV to 322 MeV,
%\footnote{\bf 1, 2, 415, 350 $->$ 1.2, 2.1, 370, 322.
%In this calculation $V_N/3$
%is fixed at 90 MeV and effective kaon mass is obtained
%by (\ref{mlow}). And the densities are obtained from the
%distribution function of free nucleons.}
the sum $\mu_B+m^\star_{\tiny{K^-}}$ decreasing very slightly
from 1230 MeV to 1227 MeV. Indeed, even at $\rho=\rho_0/4$,
$\mu_B+m^\star_{\tiny{K^-}}\sim 1215$ MeV, not much smaller.
%In the physical situation the
%temperature will increase slightly, so that even this small
%decrease will probably be over-compensated for by an increase in
%$e^{-m_\Lambda/T}$.
%The temperature does not change rapidly
%because the entropy density in compression is increased more by
%production of low-mass particles than by increasing the
%temperature.

We believe that the temperature will change only little in the
region of dropping masses because in a consistent evolution
(which we do not carry out here) the scalar field energy
$m_\sigma^2\sigma^2/2$ in a mean field theory plays the role of an
effective bag constant. In ref.\cite{BBR} this is phrased in
terms of a modified Walecka theory, \be B_{eff}=\frac12
m_\sigma^2\sigma^2\Rightarrow\frac12 m_\sigma^2 (M_N/g_{\sigma
NN})^2, \ee the $\sigma$ going to $M_N/g_{\sigma NN}$ as the
nucleon effective mass goes to zero with chiral restoration. Most
of the energy with compression to higher densities goes with this
effective bag constant, estimated \cite{BBR} to be $\sim 280$
MeV/fm$^3$, rather than heat, mocking up the behavior of a mixed
phase with constant temperature. Moreover at $\rho=2\rho_0$ where
$m_N^\star$ may be $\sim 0.5 m_N$, only about 25\% of the bag
constant $B$ may have been achieved, so there may be some
increase in temperature. We shall, at the same level of accuracy,
have to replace $m_{\bar{K}}$ in the prefactor of eq.(\ref{Rp}) by
$\wkm$. We adjust the increase in temperature so that it exactly
compensates for the decrease in prefactor so that the \Kp/\Km
ratio is kept the same, as required by experiment~\cite{menzel}.
We then find that the temperature at $\rho=2\rho_0$ must be
increased from 70 to 95 MeV. This is roughly the change given by
that in inverse slopes of \Km and \Kp transverse momentum
distributions found in going from low multiplicities to high
multiplicities~\cite{menzel}.

In any case, we see that $R$ will depend but little
on density. This near cancellation of changes in the factors is
fortunate because the $K^- +p \leftrightarrow \Lambda$ reaction,
operating in the negative strangeness sector, is much stronger
than the positive strangeness reactions, so the former should
equilibrate to densities well below $2\rho_0$ and we can see that
the ``apparent equilibration" might extend all the way down to
$\rho\sim \rho_0/4$.

The near constancy of $R$ with density also explains the fact the
\Kp/\Km ratio does not vary with centrality~\footnote{We are
grateful to Helmut Oeschler for pointing this out to us.}.
Although $R$ is the ratio of $\Lambda$'s to \Km's, both of which
are in the negative strangeness sector, nonetheless, the number
of \Kp's must be equal to the sum of the two and since the
$\Lambda$'s are much more abundant than the \Km's, $R$
essentially  represents the \Kp/\Km ratio.

\begin{figure} \centerline{\epsfig{file=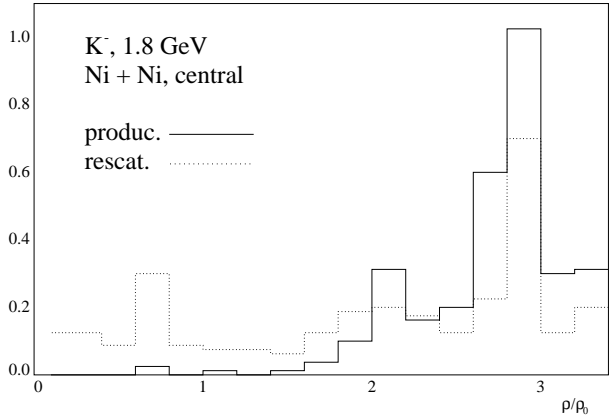,width=8cm}}
\caption{\small Calculations by Bratkovskaya and Cassing (private
communication) which show the density of origin and that of the last
interaction of the $K^-$ mesons.}\label{nini}
\end{figure}
Detailed transport results of Bratkovskaya and Cassing (see
Fig.\ref{nini}) show the last scattering of the detected \Km to be
spread over all densities from $\rho_0/2$ to $3\rho_0$, somewhat
more of the last scatterings to come from the higher density.
This seems difficult to reconcile with a scenario of the \Km
numbers being decided at one definite density and temperature,
but given our picture of dropping masses, one can see that the
\Kp/\Km ratio depends little on density ($\mu_B$) at which the
\Km last interacts. In any case we understand from our above
argument that the apparent density of equilibration can be chosen
to be very low in a thermal description and still get more or
less correct \Kp/\Km ratio.

\section{The Top-Down Scenario of $K^\pm$ Production}
\itt
 Brown and Rho~\cite{BR96} discussed fluctuations in the kaon
 sector in terms of a simple Lagrangian
 \be
 \delta {\cal
L}_{KN}=\frac{-6i}{8{f^\star}^2}(\overline{N}\gamma_0
N)\overline{K}\del_t K +
\frac{\Sigma_{KN}}{{f^\star}^2}(\overline{N}N)\overline{K}K+\cdots\equiv
{\cal L}_\omega +{\cal L}_\sigma+\cdots\label{kaonL} \ee
 It was suggested there that at high densities, the constituent
quark or quasi-quark  description can be used with the
$\omega$-meson coupling to the kaon viewed as a {\it matter field}
(rather than as a Goldstone boson). Such a description suggests
that the $\omega$ coupling to the kaon which has one non-strange
quark is 1/3 of the $\omega$ coupling to the nucleon which has
three non-strange quarks. The ${\cal L}_\omega$ in the Lagrangian
was obtained by integrating out the $\omega$-meson as in the
baryon sector. We may therefore replace it by the interaction
 \be
V_{K^\pm}\approx \pm \frac 13 V_N.
 \ee
In isospin asymmetric matter, we shall have to include also the
$\rho$-meson exchange~\cite{BR96} with the vector-meson coupling
treated in the top-down approach.

For the top-down scenario, we should replace the chiral
Lagrangian (\ref{kaonL}) by one in which the ``heavy" degrees of
freedom figure explicitly. This means that
$\frac{1}{2{f^\star}^2}$ in the first term of (\ref{kaonL})
should be replaced by ${g^\star}^2/{m_\omega^\star}^2$ and
$\frac{\Sigma_{KN}}{{f^\star}^2}$ in the second term by $\frac 23
m_K \frac{{g_\sigma^\star}^2}{{m_\sigma^\star}^2}$ assuming that
both $\omega$ and $\sigma$ are still massive. We will argue in the
next section that while the $\omega$ mass drops, the ratio
${g^\star}^2/{m_\omega^\star}^2$ should stay constant or more
likely decrease with density and that beyond certain density above
nuclear matter, the vector fields should decouple. On the other
hand, $g_\sigma$ is not scaled in the mean field that we are
working with; the motivation for this is given in Brown, Buballa
and Rho~\cite{BBR} who construct the chiral restoration
transition in the mean field in the Nambu-Jona-Lasinio model. Thus
 \be
\frac{\Sigma_{KN}}{f^2}\approx \frac 23 m_K
\frac{g_\sigma^2}{{m_\sigma^\star}^2}.\label{sigmatop}
 \ee
In this framework, $m_\sigma^\star$ is the order parameter for
chiral restoration which drops {\`a la}\, BR scaling~\cite{BR91}:
 \be
\frac{m_\sigma^\star}{m_\sigma}\equiv \Phi (\rho)\simeq
\frac{1}{1+y\rho/\rho_0}
 \ee
with $y\simeq 0.28$, at least for $\rho\lsim
\rho_0$~\footnote{$y$ may well be different from this value for
$\rho>\rho_0$. In fact the denominator of $\Phi (\rho)$ could
even be significantly modified from this linear form. At present
there is no way to calculate this quantity from first
principles.}.

\begin{figure} \centerline{\epsfig{file=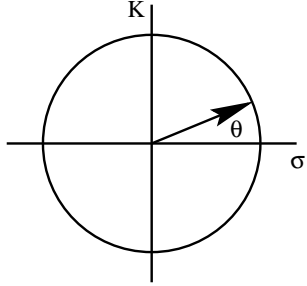,width=4cm}}
\caption{\small Projection onto the $\sigma, K$ plane. The
angular variable $\theta$ represents fluctuation toward kaon mean
field.\label{Ksigma}}
\end{figure}

Once the vector is decoupled, a simple way to calculate the
in-medium kaon effective mass, equivalent to using the
${\calL}_\sigma$, is to consider the kaon as fluctuation about
the ``$\sigma$"-axis in the V-spin formalism~\cite{BKR87} as
depicted pictorially in Fig.\ref{Ksigma}. The Hamiltonian for
explicit chiral symmetry breaking is
 \be
H_{\chi SB}&=&\Sigma_{KN}\langle\overline{N}N\rangle \cos (\theta)
+\frac 12 m_K^2 f^2 \sin^2 (\theta)\nonumber\\
&\simeq&\Sigma_{KN}\langle\overline{N}N\rangle
(1-\frac{\theta^2}{2})+ \frac 12 m_K^2 f^2 \theta^2
 \ee
where the last expression is obtained for small fluctuation
$\theta$. Dropping the term independent of $\theta$, we find
 \be
{m_K^\star}^2=m_K^2\left(1-\frac{\Sigma_{KN}\langle\overline{N}N\rangle}{f^2
m_K^2}\right).\label{mlow}
 \ee
Using eq. (\ref{sigmatop}) we obtain
 \be
{m_K^\star}^2=m_K^2\left(1-\frac 23
\frac{g_\sigma^2\langle\overline{N}N\rangle} {{m_\sigma^\star}^2
m_K}\right).\label{mhigh}
 \ee
In accord with Brown and Rho~\cite{BR96} we are proposing that
eq.(\ref{mlow}) should be used for low densities, in the
Goldstone description of the $K^\pm$, and that we should switch
over to eq.(\ref{mhigh}) for higher densities. It is possible that
the $m_K$ appearing in (\ref{mhigh}) should be replaced by
$m_K^\star$ for self-consistency but the dropping of
${m_\sigma^\star}^2$ makes the $m_K^\star$ of (\ref{mhigh})
decrease more rapidly than that of (\ref{mlow}) so that
eq.(\ref{mlow}) with $\langle\overline{N}N\rangle$ set equal to
the vector density $\rho$, a much used formula valid to linear
order in density,
 \be
{m_K^\star}^2\approx m_K^2\left(1-\frac{\rho\Sigma_{KN}}{f^2
m_K^2}\right)\label{mlowp}
 \ee
obviously gives too slow a decrease of $m_K^\star$ with density.

Although the above are our chief points, there are two further
points to remark. One, even without scaling, our vector
interaction on the kaon is still too large. Two, more
importantly, there is reason to believe in the large
$\Sigma_{KN}$ term,
 \be
\Sigma_{KN}\sim 400\ \ {\rm MeV}.\label{value}
 \ee
This comes from scaling of the pion sigma term
 \be
\Sigma_{KN}\equiv \frac{(m_u+m_s)\langle
N|\bar{u}u+\bar{s}s|N\rangle}{(m_u+m_d)\langle
N|\bar{u}u+\bar{d}d|N\rangle}\Sigma_{\pi N}.
 \ee
Taking $m_s\sim 150$ MeV, $m_u+m_d\sim 12$ MeV, $\Sigma_{\pi
N}=46$ MeV and $\langle N|\bar{s}s|N\rangle\sim \frac 13 \langle
N|\bar{d}d|N\rangle$ from lattice calculations~\cite{liu}, one
arrives at (\ref{value}).

Other authors, in adjusting the $\Sigma$ term to fit the
kaon-nucleon scattering amplitudes, have obtained a somewhat
smaller $\Sigma_{KN}$. This can be understood in the chiral
perturbation calculation of C.-H. Lee~\cite{chl} where the only
significant effect of higher chiral order terms can be summarized
in the ``range term"\footnote{In the language of heavy-baryon
chiral perturbation theory, this corresponds to the ``1/m"
correction term.}; namely $\Sigma_{KN}$ is to be replaced by an
effective $\Sigma$,
 \be
(\Sigma_{KN})_{eff}=(1-0.37\omega_K^{\star 2}/m_K^2)\Sigma_{KN}.\label{sigmaeff}
 \ee
It should be pointed out that although the $\Sigma_{KN}$ is
important at low densities,
%at which $\omega_K^\star \gsim m_K$ as the
$\omega_K$ decreases with $m_K^\star$, this ``range-term"
correction becomes less important at higher densities. This
effect -- which is easy to implement -- is included in the
realistic calculations.

\section{Partial Decoupling of the Vector Interaction}
\subsection{Evidence}\label{3.1}
\itt
 There are both theoretical and empirical reasons why we believe
 that the vector interaction should decouple at high density.
\ben
 \item
We first give the theoretical arguments. We know of three
theoretical reasons why the vector coupling $g_\omega^\star$
should drop with density.
 \bitem
 \item The first is the observation by Song et al~\cite{song}
that describing nuclear matter in terms of chiral Lagrangian in
the mean field requires the ratio $g_\omega^\star/m_\omega^\star$
to at least be roughly constant or even decreasing as a function
of density. In fact to quantitatively account for non-linear
terms in a mean-field effective Lagrangian, a dropping ratio is
definitely favored~\footnote{Since the non-linear terms -- though
treated in the mean field -- are fluctuation effects in the
effective field theory approach, this represents a quantum
correction to the BR scaling.}. For instance, as discussed in
\cite{song}, the in-medium behavior of the $\omega$-meson field is
encoded in the ``FTS1" version of the non-linear theories of
ref.\cite{FTS1}. In fact, because of the attractive quartic
$\omega$ term in the FTS1 theory, the authors of \cite{FTS1} have
(for the parameter $\eta=-1/2$ favored by experiments)
${g_\omega^\star}^2/{m_\omega^\star}^2\simeq
0.8{g_\omega}^2/{m_\omega}^2$ as modification of the quadratic
term when rewritten in our notation . In other words, their
vector mean field contains a partial decoupling already at
$\rho\approx \rho_0$ although they do not explicitly scale
$g_\omega$ as we do.

Historically, Walecka-type mean field theories with only
quadratic interactions (i.e., linear Walecka model) gave
compression moduli $K\sim 500$ MeV, about double the empirical
value. This is cured in nonlinear effective field theories like
FTS1 by higher-dimension non-renormalizable terms which
effectively decrease the growth in repulsion in density. As
suggested in \cite{song}, an effective chiral Lagrangian with BR
scaling can do the same (by the increase in magnitude of the
effective scalar field with density) but more economically and
efficiently.
\item The second reason is perhaps more theoretical. Kim and
Lee~\cite{RGKL} have shown recently that in an effective QCD
Lagrangian with baryons, pions and vector mesons put together in
hidden gauge symmetric theory, the ratio $g_V^\star/m_V^\star$
(where $V$ stands for hidden gauge bosons) falls very rapidly
with baryon chemical potential. The main agent for this behavior
is found in \cite{RGKL} to be the pionic one-loop contribution
linked to chiral symmetry which is lacking in the mean-field
treatment for BR scaling~\cite{BR91}.

One may argue on a more profound ground that the vector decoupling
in approaching chiral phase transition is a flow to a fixed
point. It has been argued recently by Harada and
Yamawaki~\cite{VM} that chiral symmetry restoration may
correspond to a ``vector manifestation" of chiral symmetry where
the octet of Goldstone pions and the octet of longitudinal vector
mesons belong to the representation $(8,1)\oplus (1,8)$. At this
point, the vector coupling goes zero \`a la Georgi's vector limit
as does the vector meson mass~\cite{georgi}.

The above arguments were made for the $\rho$ which has a simple
interpretation in terms of hidden gauge symmetry but it will
apply to the $\omega$ if the nonet symmetry continues to hold in
nuclear medium. It is difficult to be quantitative as to how fast
the ratio falls but it is clear that the drop is substantial
already near normal nuclear matter density.
 \item Finally, close
to chiral restoration in temperature, there is clear evidence from
QCD for an equally rapid drop, specifically, from the quark number
susceptibility that can be measured on the lattice~\cite{BRPR}.
The lattice calculation of the quark number susceptibility dealt
with quarks and the large drop in the (isoscalar) vector mean
field was found to be due chiefly to the change-over from hadrons
to quarks as the chiral restoration temperature is approached
from below. The factor of 9 in the ratio $g_{\omega
NN}^2/g_{\omega QQ}^2$ (where $Q$ is the constituent quark)
should disappear in the change-over.
%We show in Appendix B that
%the change-over from nucleons to quark degrees of freedom (the
%quarks being constituent quarks with residual masses) must be
%made {\it before} the chiral restoration transition. It will be
%seen that nucleons cannot make a phase transition into massless
%quarks.
Now since the electroweak properties of a constituent
quark (quasiquark) are expected to be the same as those of a bare
Dirac particle with $g_A=1$ and no anomalous magnetic moment
(i.e., the QCD quark)~\cite{weinbergquark} with possible
corrections that are suppressed as $1/N_c$~\cite{peris}, there
will be continuity between before and after the chiral
transition. This is very much in accordance with the
``Cheshire-Cat picture" developed elsewhere~\cite{NRZ}. In fact,
it is possible to give a dynamical (hadronic) interpretation of
the above scenario. For instance in the picture of \cite{KRBR},
this may be understood as the ``elementary" $\omega$ strength
moving downwards into the ``nuclear" $\omega$, the $[N^\star
(1520) N^{-1}]^{J=1,I=0}$ isobar-hole state involving a
single-quark spin flip~\cite{riskabrown}. The mechanism being
intrinsic in the change-over of the degrees of freedom, we expect
the same phenomenon to hold in density as well as in temperature.
The upshot of this line of argument is that the suppression of
the vector coupling is inevitable as density approaches the
critical density for chiral transition.
 \eitem

We believe that the different behavior of vector and scalar mean
fields, the latter to be discussed below, follows from their
different roles in QCD. With the vector this is made clearer in
the lattice calculations of the quark number susceptibility which
involves the vector interactions. In Brown and Rho~\cite{BRPR},
it is shown that as the description changes from hadronic to
quark/gluonic at $T\sim T_c$, the critical temperature for chiral
restoration, the vector interaction drops by an order of
magnitude, much faster than the logarithmic decrease due to
asymptotic freedom. We expect a similar feature in density,
somewhat like in the renormalization-group analysis for the
isovector vector meson $\rho$ of Kim and Lee~\cite{RGKL}. The
scalar interaction, on the other hand, brings about chiral
restoration and must become more and more important with
increasing density as the phase transition point is approached.

\item
{}From the empirical side, the most direct indication of the
decoupling of the vector interaction is from the baryon
flow~\cite{Bflow} which is particularly sensitive to the vector
interaction. The authors of \cite{Bflow} find a form factor of
the form
 \be
 f_V (\Bp)= \frac{\Lambda_V^2-\frac 16 \Bp^2}{\Lambda_V^2+\Bp^2}
 \ee
with $\Lambda_V=$0.9 GeV is required to understand the baryon
flow. Connecting momenta with distances, one finds that this
represents a cutoff at
 \be
R_{cutoff}\sim \frac{\sqrt{6}}{\Lambda_V}\sim 0.5\ \ {\rm fm}.
 \ee

Furthermore it is well known that the vector mean field of the
Walecka model must be modified, its increase as $E/m_N$ removed,
at a scale of $\sim 1$ GeV. The reason for this is presumably that
inside of $R\sim 0.5$ fm, the finite size of the solitonic
nucleon must be taken into account. A repulsion still results,
but it is scalar in nature as found in \cite{vento} and for which
there are direct physical indications~\cite{holinde}.
 \een
\subsection{Kaons at GSI}
\itt
 The \Kp and \Km energies in the top-down scenario are given
by
 \be
\omega_{\pm}=\pm \frac 13 \frac{\omega_{\pm}}{m_K}V_N
+\sqrt{k^2+{m_K^\star}^2}.\label{energy}
 \ee
Although at high densities it will decouple, the term linear in
$V_N$ that figures in the range correction in
$(\Sigma_{KN})_{eff}$ will give a slightly different effective
mass for \Kp and \Km before decoupling. Although the large
distance vector mean field must arise from vector meson exchange,
this must be cut off at a reasonably large distance, say, $\sim$
0.5 fm as indicated by the baryon-flow mentioned above.

For the GSI experiments with temperature $\sim 75$ MeV, the
momentum is $|\Bp|\sim m_N/2$, and
 \be
{f_V}^2(p)\sim 0.6.
 \ee
We therefore propose to use
\be
 \frac{V_K}{m_K}=\frac 38 (\frac{f_V^2}{m_K})^2
 \frac{\rho}{m_K}\approx 0.07.
 \ee
This is small. Furthermore we assume it to be constant above
$\rho_0$.

The above arguments could be quantified by a specific model. For
example, as alluded to above, the low-lying $\rho$- and
$\omega$-excitations in the bottom-up model can be built up as
$N^\star$-hole excitations~\cite{KRBR}. At higher densities,
these provide the low-mass strength. One might attempt to
calculate the coupling constants to these excitations in the
constituent-quark (or quasi-quark) model, which as we have
suggested would be expected to be more applicable at densities
near chiral restoration. Riska and Brown~\cite{riskabrown} find
the quark model couplings to be a factor $\sim 2$ lower than the
hadronic ones~\cite{frimanetal}.

\section{Schematic Model}
\subsection{First try}
\itt
 On the basis of our above considerations, a first try in
transport calculation might use the vector potential with the
Song scaling~\cite{song} as
$g_\omega^\star/m_\omega^\star=constant$ and the effective mass
 \be
{m_K^\star}^2\approx
m_K^2\left(1-\frac{\rho(\Sigma_{KN})_{eff}}{f^2
m_K^2}\right)\label{mlowpp}
 \ee
with $\Sigma_{KN}=400$ MeV and $(\Sigma_{KN})_{eff}$ given by eq.
(\ref{sigmaeff}). While as argued above the vector coupling will
decouple at very high densities, as $\omega_K$ drops, the vector
potential will become less important even at moderate densities
since the factor $\omega_K/m_K$ comes into the coupling of the
vector potential to the kaon.

Our schematic model (\ref{mlowpp}) gives roughly the same mass as
used by Li and Brown~\cite{LB} to predict kaon and antikaon
subthreshold production at GSI. For $\rho\sim 3\rho_0$ it gives
$m_{K^-}^\star\sim 230$ MeV, less than half the bare mass. We
predict somewhat fewer $K^-$-mesons because the (attractive)
vector interaction is largely reduced if not decoupled. Cassing et
al~\cite{cassing} have employed an $m_{K^-}^\star$ somewhat lower
than that given by eq.(\ref{mlowpp}) to describe a lot of data.

The experimental data verify that our description is quite good
up to the densities probed, i.e., $\sim 3\ \rho_0$. In order to
go higher in density, we switch to our top-down description which
through eq.(\ref{sigmatop}) involves
$g_\sigma^2/{m_\sigma^\star}^2$. Although the scalar interaction
could have roughly the same form factor as the vector, cutting it
off at $\sim 0.5$ fm as mentioned above,  we believe that this
will be countered by the dropping scalar mass $m_\sigma^\star$
which must go to zero at chiral restoration (viewed as an order
parameter). Treating the scalar interaction linearly as a
fluctuation (as in (\ref{sigmatop})) cannot be expected to be
valid all the way to chiral restoration but approaching the
latter the $\sigma$-particle becomes the ``dilaton" in the sense
of Weinberg's ``mended symmetry"~\cite{weinberg-salam,beane} with
mass going to zero (in the chiral limit) together with the pion .

At high densities at which the vector interaction decouples, the
$K^+$ and $K^-$ will experience nearly the same very strong
attractive interactions. This can be minimally expressed through
the effective mass $m_K^\star$. At low densities where the vector
potential not only comes into play but slightly predominates over
the scalar potential, the $K^+$ will have a small repulsive
interaction with nucleons. It is this interaction, extrapolated
without medium effects by Bratkovskaya et al~\cite{brat} which
gives the long equilibration time of 40 fm/c. However, clearly
the medium effects will change this by an order of magnitude.
\subsection{Implication on kaon condensation and maximum
neutron-star mass}
 \itt
While in heavy-ion processes, we expect that taking
$m_\sigma^\star$ to zero (or nearly zero in the real world) is a
relevant limiting process, we do not have to take
$m_\sigma^\star$ to zero for kaon condensation in neutron stars,
since the  $K^-$-mass $m_K^\star$ must be brought down only to
the electron chemical potential $\mu_e\simeq E_F (e)$, the
approximate equality holding because the electrons are highly
degenerate. We should mention that it has been suggested that the
electron chemical potential $\mu_e$ could be kept low by
replacing electrons plus neutrons by $\Sigma^-$ hyperons (or more
generally by exploiting Pauli exclusion principle with hyperon
introduction) in neutron stars~\cite{glendenning}. In this case,
the $\mu_e$ might never meet $m_K^\star$.
%However, up to nuclear matter density
%$\rho_0$ the $\Sigma^-$-nuclear interaction has been shown to be
%repulsive~\cite{batty} and it would have to be attractive at
%higher densities in order to bring $\mu_e$ down. We have no data
%to either support or refute this, whereas we do have data to show
%that the $K^-$ experiences strong attraction at high densities.
%Nor do we know of any plausible theoretical mechanisms that might
%turn the repulsion into an attraction above $\rho_0$. On the
%contrary, we believe that the decoupling of the vector mean field
%at higher densities will continue to disfavor the {\it naive}
%replacement of neutron and electron by $\Sigma^-$ or neutron by
%$\Lambda$.

Hyperon introduction may or may not take place, but even if it
does, the scenario will be more subtle than considered presently.
To see what can happen, let us consider what one could expect
from a naive extrapolation to the relevant density, i.e.,
$\rho\sim 3\rho_0$, based on the {\it best} available nuclear
physics. The replacement of neutron plus electron will take place
if the vector mean field felt by the neutron is still high at that
density. The threshold for that would be
 \be
E_F^n +V_N +\mu_e\simeq M_{\Sigma^-} + \frac 23 V_N
+S_{\Sigma^-}\label{thr}
 \ee
where $E_F^n$ is the Fermi energy of the neutron, $M_{\Sigma^-}$
the bare mass of the $\Sigma^-$ and the $S_{\Sigma^-}$ the scalar
potential energy felt by the $\Sigma^-$. Here we are simply
assuming that the two non-strange quarks of the ${\Sigma^-}$
experience $2/3$ of the vector mean field felt by the neutron.
Extrapolating the FTS1 theory~\cite{FTS1}\footnote{There is
nothing that would suggest that the effective Lagrangian valid up
to $\rho\sim \rho_0$ will continue to be valid at $\rho\sim
3\rho_0$ without addition of higher mass-dimension operators,
particularly if the chiral critical point is nearby. So this
exercise can be taken only as indicative.}
and taking into
account in $V_N$ the effect of the $\rho$-meson using vector
dominance , we find $E_F^n+V_N\sim 1064$ MeV
at $\rho\approx 3\rho_0$. From the
extended BPAL 32 equation of state with compression modulus 240
MeV~\cite{prakash}, the electron chemical potential comes out to
be $\mu_e\simeq 214$ MeV. So the left-hand side of (\ref{thr}) is
$E_F^n +V_N +\mu_e\sim 1258$ MeV. For the right-hand side, we use
the scalar potential energy for the $\Sigma^-$ at $\rho\approx
3\rho_0$ estimated by Brown, Lee and Rapp~\cite{BLR} to find that
$M_{\Sigma^-} + \frac 23 V_N +S_{\Sigma^-}\sim 1240$ MeV. The
replacement of neutron plus electron by $\Sigma^-$ looks favored
but only slightly.

What is the possible scenario on the maximum neutron star mass if
we continue assuming that the calculation we made here can be
trusted?  A plausible scenario would be as follows. \Km
condensation supposedly occurs at about the same density and both
the hyperonic excitation (in the form of $\Sigma N^{-1}$ -- where
$N^{-1}$ stands for the nucleon hole -- component of the
``kaesobar"~\cite{BLR}) and \Km condensation would occur at
$T\sim 50$ MeV relevant to the neutron-star matter. Now if as is
likely the temperature is greater than the difference in energies
between the two possible phases, although the hyperons will be
more important initially than the kaons, all of the phases will
enter more or less equally in constructing the free energy of the
system. In going to higher density the distribution between the
different phases will change in order to minimize the free
energy. Then it is clear that dropping from one minimum to
another, the derivative of the free energy with density -- which
is just the pressure -- will decrease as compared with the
pressure from any single phase. This would imply that the maximum
neutron star mass calculated with either hyperonic excitation or
kaon condensation alone must be greater than the neutron star
mass calculated with inclusion of both.

The story will be quite different if the vector field decouples.
We showed in Section \ref{3.1} that the isoscalar vector mean
field must drop by a factor $\gsim 9$ in the change-over from
nucleons to quasiquarks as variables as one approaches the chiral
restoration density. Hyperons will disappear during this drop. It
is then inevitable that the kaon will condense {\it before} chiral
restoration and that the kaon condensed phase will persist
through the relevant range of densities which determine the
maximum neutron star mass.

\section{Concluding Remarks}
 \itt
By now there is a general consensus that the light-quark hadrons
must behave differently in medium than in free space. This is
understood in terms of a vacuum change induced by medium {\it \`a
la}\, QCD. In this paper, we are re-confirming this property by
arguing that not only the kaon mass but also its coupling to
vector mesons should drop in matter with density. In particular,
with the introduction of medium effects the apparent equilibration
found in strangeness production at GSI can be increased from the
baryon number density of $\sim\frac 14\rho_0$ up to the much more
reasonable $\sim 2\rho_0$.

{}From the baryon flow analysis we have direct indications that the
 vector interaction decouples from the nucleon at a three-momentum
 of $|\Bp|\sim 0.9$ GeV/c or at roughly 0.5 $m_N c$. In colliding
 heavy ions this is reached at a kinetic energy per nucleon of
 $\sim \frac 18 m_N c^2$ which means a temperature of 78 MeV when
 equated to $\frac 32 T$. This is just the temperature for
chemical freeze-out at GSI energies. We have given several
theoretical arguments why the vector coupling should drop rapidly
with density.

Once the vector mean field, which acts with opposite signs on the
$K^+$ and $K^-$ mesons is decoupled, these mesons will feel the
same highly attractive scalar meson field. Their masses will fall
down sharply; e.g., from eq.(\ref{mlowpp}) with proposed
parameters,
 \be
 \frac{m_K^\star}{m_K}\sim 0.5
 \ee
at $\rho\approx 3\rho_0$ and possibly further because of the
dropping $m_\sigma^\star$. The differing slopes of $K^+$ and
$K^-$  with kinetic energy will then develop after chemical
freeze-out, as suggested by Li and Brown~\cite{LB}.

In this paper we have focused on the phenomenon at GSI energies.
Here the chief role that the dropping \Km-mass played was to keep
the combination $\mu_B+ \wkm$ nearly constant so
that low freeze-out density in the thermal equilibration scenario
became irrelevant for the \Kp/\Km ratio. We suggest that the same
scenario applies to AGS physics, where the freeze-out density in
the thermal equilibration picture comes out to be $\sim
0.35\rho_0$~\cite{peter}.
In increasing this density to $\sim
%0.6\rho_0$,
0.57\rho_0$, which we consider more reasonable, $\mu_B+ \wkm$
changes from
%990 MeV only to 1011 MeV, an $\sim 2$
992 MeV only to 1026 MeV, an $\sim 3$ \% change, when only the
density dependence is put into $\wkm$. It is likely to change
less when the temperature dependence, that we are now working on,
is added. In fact, there is no discernible dependence on
centrality in the \Km/\Kp ratios measured at $4A$ GeV, $6A$ GeV,
$8A$ GeV and $10.8A$ GeV~\cite{dunlop}. From this it follows
either that the ratio of produced \Km to hyperons is nearly
independent of density or that the negative strangeness
equilibrates down to a lower freeze-out density and then
disperses. Given the relative weakness of the strange
interactions, we believe the former to be the case. In fact, we
suggest that near constancy with multiplicity of the \Km/\Kp
ratio found experimentally be used to determine temperature
dependence of $\wkm$ in the region of temperatures reached in AGS
physics. As was done for GSI energies, the temperature can be
obtained from the inverse slopes of the kaon and antikaon
distributions of $p_\perp$ and then the temperature dependence of
$\wkm$ can be added to the density dependence as in \cite{BKWX},
in such a way that $\mu_B+\wkm$ stays roughly constant as
function of density. At least this can be done in the low-density
regime considered in \cite{peter} when the approximation of a
Boltzmann gas is accurate enough to calculate $\mu_B$. Our
``broad-band equilibration"; i.e., the production of the same,
apparently equilibrated ratio of \Km -mesons to hyperons over a
broad band of densities, avoids complications in the way in which
the \Km\ degree of freedom is mixed into other degrees of freedom
at low density~\cite{Ramos:2000ku}. Most of the \Km -production
will take place at the higher densities, as shown in Fig.
\ref{nini}, where the degrees of freedom other than \Km\ have
been sent up to higher energies by the Pauli principle.

Unless the electron chemical potential in dense neutron star
matter is prevented from increasing with density (as might happen
if the repulsive $\Sigma$-nuclear interaction turns to attraction
at $\rho> \rho_0$), kaon condensation will take place {\it before}
chiral restoration. This has several implications at and beyond
chiral restoration. For instance, its presence would have
influence on the conjectured color superconductivity at high
density, in particular regarding its possible coexistence with
Overhauser effects, skyrmion crystal and other phases with
interesting effects on neutron star cooling.

The phenomenon of vector decoupling, if confirmed to be correct,
will have several important spin-off consequences. The first is
that it will provide a refutation of the recent
claim~\cite{pandh} that in the mean-field theory with a
kaon-nuclear potential given by the vector-exchange
(Weinberg-Tomozawa) term -- both argued to be valid at high
density -- kaon condensation would be pushed up to a much higher
density than that relevant in neutron-star matter. Our chief point
against that argument is that the vector decoupling and the
different role of scalar fields in QCD (e.g., BR scaling)
described in this article cannot be accessed by the mean field
reasoning used in \cite{pandh} or by any other standard nuclear
physics potential models. The second consequence that could be of
a potential importance to the interpretation of heavy-ion
experiments is that if the vector coupling~\footnote{While the
decoupling of the isoscalar vector interaction near the chiral
restoration is highly plausible by the disappearance of a factor
of 9 in the change-over from hadrons to quarks, the decoupling of
the isovector vector interaction seems to be a lot more complex
as indicated by the RG analysis of Kim and Lee~\cite{RGKL}. What
is not difficult to see is that their decoupling should occur at
the same point as indicated in the quark number
susceptibility~\cite{BRPR}.} rapidly diminishes with density, the
strong-coupling perturbation calculation of the vector response
functions used in terms of ``melting" vector mesons to
explain~\cite{rapp}, e.g., the CERES dilepton data must break
down as one approaches the decoupling point that is out of reach
of strong-coupling perturbative methods. It in turn provides yet
one more justification (in addition to what has been already
offered~\cite{migdal}) to the quasi-particle description in dense
matter encoded in BR scaling and exploited in this paper.

Finally we should stress that the scenario presented in this
paper -- which is anchored on Brown-Rho scaling-- should not be
considered as an {\it alternative} to  a possible quark-gluon
scenario currently favored by the heavy-ion community. It is more
likely a sort of ``dual" description of the same physics along
the line that the Cheshire-Cat Principle~\cite{NRZ} embodies that
would continue to apply at higher energies.

\setcounter{equation}{0} \setcounter{figure}{0}
\renewcommand{\theequation}{\mbox{A.\arabic{equation}}}
\renewcommand{\thefigure}{\mbox{A.\arabic{figure}}}

 \subsection*{Acknowledgments}
\itt The initial part of this work was done while we were
visiting Korea Institute for Advanced Study whose hospitality is
acknowledged. We are grateful for comments from Bengt Friman and
Chang-Hwan Lee. One of the authors (GEB) is indebted to Elena
Bratkovskaya, Peter Braun-Munzinger,
Wolfgang Cassing, Che Ming Ko, Helmut Oeschler and
Krzystof Redlich for useful discussion and criticism. GEB and CS were
supported by the US Department of Energy under Grant No.
DE-FG02-88 ER40388. The work of CS was partly supported by
Korea Science and Engineering Foundation.

\section*{Appendix A}
\itt In this appendix we show how our argument that gives a
correct $K^+/K^-$ ratio can reproduce the $K^+/\pi^+$ ratio. Let
us leave $T=70$ MeV and choose $\rho\sim 2\rho_0$ as educated
guesses. We are thereby increasing the equilibration density by a
factor $\sim 8$. We then calculate the baryon density for this
$\mu_B$ and $T=70$ MeV and find $\rho=2\rho_0$ which checks the
consistency.

 According to Brown et al~\cite{BKWX} the $K^+$ production under
 these conditions will come chiefly from $BB\rightarrow N\Lambda
 K$, with excited baryon states giving most of the production.
 {}From the solid curve for $\rho/\rho_0=2$ in fig.9 of \cite{BKWX}
 we find $\langle \sigma v\rangle\sim 2\times 10^{-3}\ {\rm mb}=
 2\times 10^{-4}\ {\rm fm}^2$. The rate equation reads
  \be
\delta \Psi_K=\frac 12 (\sigma_{BB}^{BYK} v_{BB}) n_B^2\simeq
\frac 12 (2\times 10^{-4}\ {\rm fm}^2)\rho_B^2=dn/dt=\frac
19\times 10^{-4}\ {\rm fm}^{-4}\label{A1}
 \ee
where $B$, $Y$ and $K$ stand, respectively, for baryon, hyperon
and kaon. For this, we have taken $\langle \sigma_{BB}^{BYK}
v_{BB}\rangle$ from fig.9 of \cite{BKWX} and $\rho=2\rho_0=\frac
13\ {\rm fm}^{-3}$. Choosing a time $t=10\ {\rm fm}/c$ we obtain
 \be
n_{\tiny{K^+}}\simeq\delta \Psi_K\ t= \frac 29\times 10^{-3}\ {\rm
fm}^{-3}.\label{A2}
 \ee
Now equilibrated pions have a density
 \be
n_\pi=0.37 (T/197{\rm MeV})^3\ {\rm fm}^{-3}=0.016\ {\rm
fm}^{-3}\label{A3}
 \ee
for $T=70$ MeV. From (\ref{A2}) and (\ref{A3}) we get
 \be
n_{\tiny{K^+}}/n_{\pi^+}\simeq 0.0069\label{A4}
 \ee
 which is slightly below the ``equilibrated value" of 0.0084 of Table 1 of
 Cleymans et al~\cite{cleymans2}. Production of \Kp by pions may
 increase our number by $\sim 25$\%~\cite{ko2}.

% We have learned from this simple exercise that a large increase
% in $K^+$ production from medium effects is necessary to get into
% the ballpark of ``equilibration." This is a factor of several
% increase from medium effects in going from the solid curve for
% $\rho/\rho_0=0.25$ to that for $\rho/\rho_0=2$ in fig.9 of
% \cite{BKWX}. Note, however, that these curves were calculated for
% an $E_{\tiny{K^+}}^\star$ with slightly repulsive medium effects.
% As argued in the main text, we now have an $E_{\tiny{K^+}}^\star$
% that should decrease with density, at least at the higher
% densities.
%
% For $\rho=2\rho_0$ our schematic formula gives $m_K^\star=375$
%MeV and adding the vector potential (\ref{energy}) of $V_K\simeq
%35$ MeV we
% have $E_{\tiny{K^+}}^\star=410$ MeV. We approximately correct for
% this decreased mass by changing the energy in the Boltzmann
% factor from $E_{\tiny{K^+}}+E_\Lambda-\mu_B$ to
% $E_{\tiny{K^+}}^\star+E_\Lambda-\mu_B$, assuming $\mu_B$ to be
% that given in \cite{BKWX} which is not far from $m_N c^2$. This
% gives us a correction factor of
% $exp(E_{\tiny{K^+}}-E_{\tiny{K^+}}^\star)/T\simeq
% e^{85/70}\approx 3.4$. Such a factor would bring (\ref{A4})
% closer to the ``equilibrated" ratio.
%
% It is clear from the above discussion that we can play a dropping
%$m_{\tiny{K^+}}^\star$ off  against $\mu_B$ by the amount that
%$m_{\tiny{K^+}}^\star$ decreases, just as in getting the
%empirical $K^+/K^-$ ratio.

Our discussion of \Kp production in this Appendix is in general
agreement with earlier works by Ko and
collaborators~\cite{ko2,ko1}. In fact, if applied at the quark
level, the vector mean field is conserved through the production
process in heavy-ion collisions, so it affects only the
strangeness condensation in which there is time for strangeness
non-conservation. These earlier works establish that at the GSI
energies the \Kp content remains roughly constant once the
fireball has expanded to $\sim 2\rho_0$, so that in this sense
one can consider this as a chemical freeze-out density.

It should be noted that in the papers \cite{cassing,BKWX,ko2,ko1}
and others, the net potential -- scalar plus vector -- on the
\Kp-meson is slightly attractive at $\rho\sim 2\rho_0$ even
though the repulsive vector interaction is not decoupled. Since
in our top-down description the vector interaction can be thought
of as applied to the quark (matter) field in the \Kp, the total
vector field on the initial components of a collision is then the
same as on the final ones, so the vector mean fields have
effectively no effect on the threshold energy of that process.
%We
%should remark that in \cite{tonythomas}, a vector field is strong
%enough so that the total potential on the \Kp is $\sim 50$ MeV
%repulsive at $\rho=2\rho_0$. One can see from figs.6-10 of this
%paper that threshold production of positive strangeness is
%suppressed compared with experiment.
Our proposal in this paper is that the vector mean field on the
\Kp should be below the values used by the workers in
\cite{cassing,BKWX,ko2,ko1} due to the decoupling. However, in
comparison with \cite{cassing}, our total potential on the \Kp at
$\rho=2\rho_0$ is $\sim -85$ MeV, as compared with their $\sim
-30$ MeV. We have not redone the calculation of \cite{BKWX} to
take into account this difference. \setcounter{equation}{0}
\setcounter{figure}{0}
\renewcommand{\theequation}{\mbox{B.\arabic{equation}}}
\renewcommand{\thefigure}{\mbox{B.\arabic{figure}}}

\end{document}